\documentclass{article}
\usepackage{spconf,amsmath,graphicx}
\usepackage{enumitem}
\usepackage{hyperref}
\usepackage{authblk}
\usepackage{graphicx}
\usepackage{subcaption}
\usepackage{caption}
\setlist{nosep, leftmargin=14pt}
\usepackage{mwe} 
\usepackage{amsmath}

\title{PathoGen-X: A Cross-Modal Genomic Feature Trans-Align Network for Enhanced Survival Prediction from Histopathology Images}

\name{ Akhila Krishna$^{1}$ \qquad
    Nikhil Cherian Kurian$^{2}$ \qquad
    Abhijeet Patil$^{1}$ \qquad
    Amruta Parulekar $^{1}$\qquad \\
    Amit Sethi$^{1}$
}

\address{
    $^{1}$Indian Institute of Technology Bombay, Mumbai, India \\
    $^{2}$ Australian Institute for Machine Learning, Adelaide, Australia
}

\begin{document}
\maketitle

\begin{abstract}
Accurate survival prediction is essential for personalized cancer treatment. However, genomic data -- often a more powerful predictor than pathology data -- is costly and inaccessible. We present the cross-modal genomic feature translation and alignment network for enhanced survival prediction from histopathology images (PathoGen-X). It is a deep learning framework that leverages both genomic and imaging data during training, relying solely on imaging data at testing. PathoGen-X employs transformer-based networks to align and translate image features into the genomic feature space, enhancing weaker imaging signals with stronger genomic signals. Unlike other methods, PathoGen-X translates and aligns features without projecting them to a shared latent space and requires fewer paired samples. Evaluated on TCGA-BRCA, TCGA-LUAD, and TCGA-GBM datasets, PathoGen-X demonstrates strong survival prediction performance, emphasizing the potential of enriched imaging models for accessible cancer prognosis.
\end{abstract}

\vspace{6pt}
\section{Introduction}
\label{sec:intro}
Advances in cancer survival analysis are essential for assessing progression risks and guiding treatment strategies \cite{cherian20212021,verghese2023multiscale}. Clinically, prognostic models rely on a range of biomarkers, such as imaging data and genetic profiles, to estimate outcomes \cite{ding2023pathology}. With the advent of deep learning (DL), there is growing interest in using routinely stained histopathology images to predict survival outcomes \cite{transmil, attentionmil,clam}. High-resolution whole-slide images (WSIs) capture cellular structures linked to cancer stage and severity, offering weak supervisory signals that can predict treatment outcomes. However, imaging-based markers lack the depth and specificity of molecular profiling, such as RNA sequencing, which provides more robust information on tumor progression \cite{ding2023pathology}. Though the gene-expression data are highly informative for survival predictions, their limited availability, due to extraction costs, often restricts their use \cite{cherian20212021}. Efficient utilization of available gene-expression data with the histopathology imaging data can thus significantly improve the performance of survival prediction models \cite{zhou2023cross, ding2023pathology}.

\vspace{6pt}

In this context, a common approach to improve survival predictions involves multi-modal information fusion of genomic and imaging data \cite{zhou2023cross}. However, this method usually requires both data types during training and testing, which is often impractical . A more feasible framework is to utilize genomic and imaging data during training but only requires imaging data at testing. To this end, we introduce a cross-modal genomic feature translation and alignment network (PathoGen-X), a DL pipeline for inter-modal feature alignment and translation based on the principles of “stronger signals enhancing weaker signals”. PathoGen-X consists of a stack of transformer-based encoder-decoder network that are optimized to effectively align and translate histopathology image based weaker modality signals to the feature space of RNA-seq data based strong modality feature space in the training phase. Our approach differs from popular alternative methods, such as unsupervised pre-training based on contrastive learning or similarity learning methods in which a similarity loss is optimized to project both modalities to a shared latent space \cite{ ding2023pathology}. Additionally, PathoGen-X requires fewer training pairs of samples from both modalities, making it more sample efficient than the methods that employs contrastive loss functions \cite{zhou2023cross,ding2023pathology}. In the test phase, PathoGen-X utilizes the translated projection space to achieve robust survival predictions based solely on imaging data.
\vspace{6pt}

To demonstrate the effectiveness of the proposed DL pipeline, we evaluated its performance on three publicly available datasets \cite{tcga}—TCGA-BRCA, TCGA-LUAD, and TCGA-GBM, that provide both WSI imaging and RNA-sequencing data. We show the predictive power of our methodology by comparing against popular  baselines, that employs imaging, imaging and genomic data using the concordance index (c-index) for survival prediction. Our method achieves performance comparable to a model trained solely on genomic data \emph{while using only imaging data at test time}. The superior performance of PathoGen-X, highlight the need to use any available complementary information in training DL models, such as RNA data, to work with imaging data.


\begin{figure*}[h]
    \centering
    \includegraphics[width=0.79\linewidth]{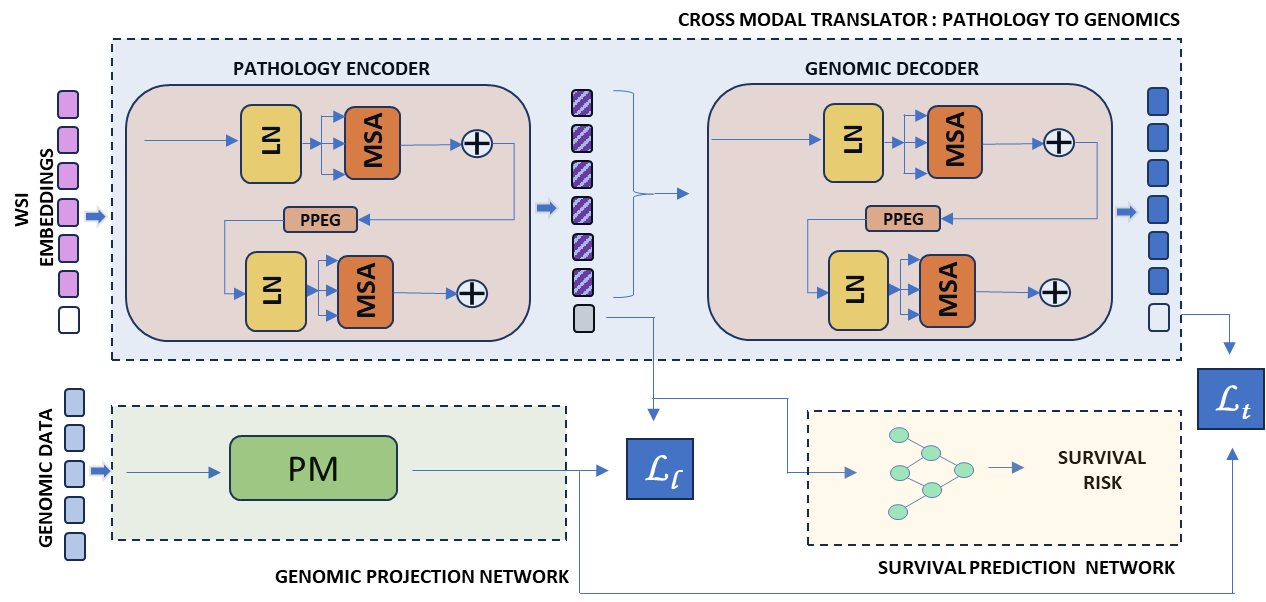}
    \caption{The cross-modal genomic feature trans-align network (PathoGen-X) features four main components: a pathology encoder, a genomic decoder, a genomic projection matrix (PM)  and a survival prediction module.} 
    \vspace{-7pt}
    
    \label{fig:main_diag}
\end{figure*}

\section{Related Work}
\vspace{-1mm}
Popular survival analysis methods \cite{transmil, attentionmil,clam} have primarily relied on clinical data and pathology. Yet, these often fall short of achieving the accuracy and reliability needed in clinical practice. As an alternative, multi-modal survival analysis models were introduced that integrate genomic data with pathology and clinical information \cite{cherian20212021}. Most of the reported studies that integrated whole slide images (WSIs) and genomic data use simple fusion techniques, such as concatenation, dot product, or attention mechanisms \cite{mobadersany2018predicting, zheng2022multi}. Although these methods combine pathology and genomic data, they often overlook deeper, intrinsic correlations and interactions between these modalities, limiting their predictive accuracy. More recent set of methods address these limitations by introducing cross-modal interaction mechanisms, as seen in \cite{zhou2023cross, long2024mugi}. However, these approaches typically require both pathology and genomic data for survival prediction, which can be challenging in clinical settings due to the high costs and logistical complexities of genomic testing compared to the relative accessibility of pathology imaging.

Additionally, existing models that rely solely on pathology images \cite{ding2023pathology} translate both modalities into a shared representation space. This approach may inadvertently incorporate redundant information that does not directly contribute to survival prediction and often demands large, diverse datasets with extensive genomic information for effective training.
\vspace{-2mm}
\vspace{6pt}
\section{Methodology}
\label{sec:methods}


Our proposed approach utilizes a transformer-based encoder-decoder architecture to translate features from pathology images into a prognostically discriminative representation space of gene expression data, for enhancing survival prediction. This deep learning pipeline is composed of three key modules: a cross modality translator to translate pathology embeddings to genomic space, a genomic feature projection network, and a task-specific survival prediction network (shown in \ref{fig:main_diag}).

\vspace{-2mm}
\subsection{Cross-Modality Translator: Pathology to Genomics}
To enable effective cross-modal alignment between pathology images and genomic data, our model incorporates a cross-modality translator that maps image-derived features into a genomic-compatible latent space. This translation is achieved through a pathology encoder $P_E$ and a genomic decoder $G_D$.
\vspace{-3mm}
\subsubsection*{Pathology Encoder ($P_{E}$)}
The pathology encoder processes pathology images and extracts relevant features that are crucial for survival prediction. We adopted a transformer-based architecture with multi-head self-attention\cite{attention} (MSA) layers to capture both local and global features from the images. The pathology embeddings, represented as \( P_0 \), are passed through the encoder, which consists of multiple MSA blocks and a positional patch embedding generation (PPEG) layer\cite{transmil}. A learnable class token is introduced to aggregate the information from all patches within a pathology image, resulting in a latent embedding that is used in subsequent alignment steps and to predict survival risk. The final output of the pathology encoder, \( P_l \), has dimensions \( (N+1) \times D \), where \( N \) is the number of patches and \( D \) is the embedding dimension.
$$P^1 = LN(MSA(P_0))+ P_0$$
\vspace{-6mm}
$$ P_2 = PPEG(P_1 )$$
\vspace{-5mm}
$$P_l = LN(MSA(P_2))+ P_2$$

\begin{figure*}[h]
    \centering
    \includegraphics[width=0.85\linewidth]{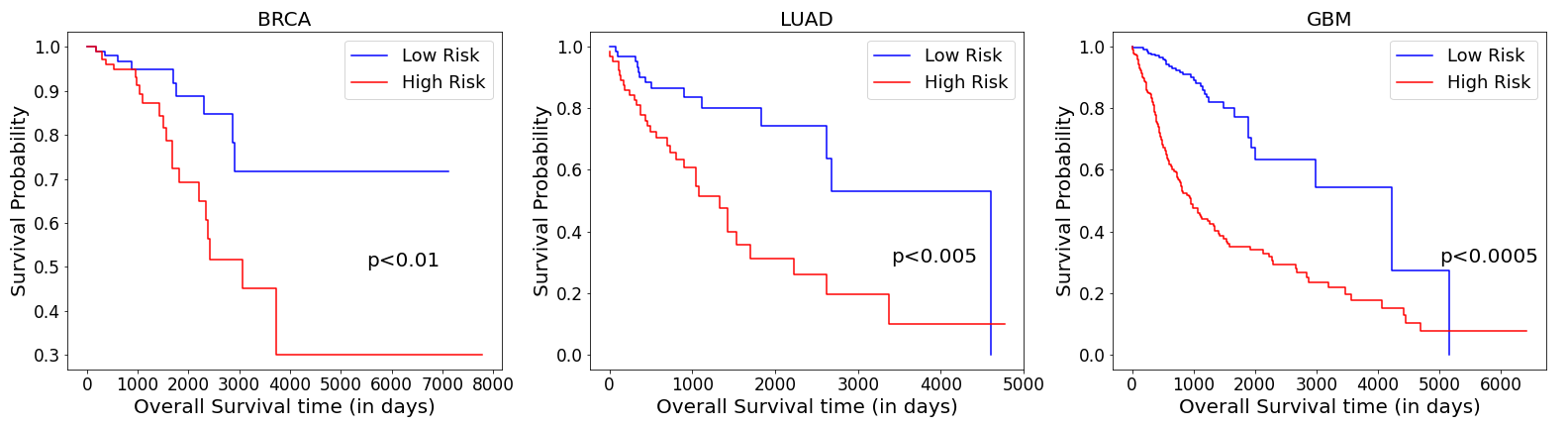}
    \caption{ The Kaplan-Meier curves of our model on the three datasets. We used median risk to stratify patients into low and high risk groups.}
    \label{fig:surv_curves}
    \vspace{-14pt}
\end{figure*}
\vspace{-4pt}
\subsubsection*{Genomic Decoder $(G_D)$}

Once both pathology and genomic data are embedded into the latent space, the outputs from the pathology encoder $(P_l)$ are passed through a genomic decoder to ensure that the features learned from pathology images can be effectively translated back into genomic representations, $G_l$. 
Architecturally, the genomic decoder is similar to the pathology encoder, utilizing MSA layers and position encoding to reconstruct genomic features from the latent space.
\vspace{-8pt}

\subsection{Genomic Feature Projection Network ($G_E$)}
The genomic feature projection network is a learnable projection matrix used to translate genomic embeddings ($G_0)$ to latent representation ($G_l$) that align with the latent representation embeddings from the pathology-to-latent encoder, $P_E$. 
\vspace{-5pt}
$$G_l = G_E(G_0)$$

\vspace{-8pt}

\subsection{Survival prediction network}
To predict survival outcomes, we add a multi-layer perceptron (MLP) layer on top of the latent space representation, which is responsible for predicting the risk of death. We use Cox loss function \cite{cox} in survival analysis to optimize the neural network parameters to estimate survival risk. The Cox loss function, \( L_{\text{Cox}} \), is defined as:
\vspace{-2pt}
\[
L_{\text{Cox}} = - \sum_{i \in \text{events}} \left( f(X_i) - \log \sum_{j \in R(T_i)} \exp(f(X_j)) \right),
\]

\vspace{2pt}
where \( f(X_i) \) represents the risk score for individual \( i \) output by the neural network, \( T_i \) is the observed survival time for individual \( i \), and \( R(T_i) \) denotes the risk set, which includes individuals who are still at risk at time \( T_i \). 

\vspace{-4pt}
\subsection{Cross Modal Alignment and Translation}

To ensure that the features learned from both pathology images and genomic data are aligned, we introduce two loss functions: the latent loss and the translation loss.

\begin{itemize}
    \item \textbf{Latent Loss (\( L_l \))}: This loss function ensures that the latent embeddings from the pathology and genomic encoders are similar. The latent loss is weighted sum of Euclidean distance and KL divergence between the outputs of the pathology encoder (\( P_l \)) and the genomic encoder (\( G_l \)):
    \vspace{-6mm}

    \[
        D_{KL}(P_l \parallel G_l) = \sum_i P_l(i) \log \frac{P_l(i)}{G_l(i)}
        \]
        \vspace{-4mm}
        \[
        L_l = \lambda_1 D_{KL}(P_l, G_l) + \lambda_2 \|P_l - G_l\|^2
  \]
        By minimizing this loss, we ensure cross-modal alignment in the latent space.
    \item \textbf{Translation Loss (\( L_t \))}: The translation loss ensures that the pathology embeddings can be accurately translated into genomic projection ($G_l$) via the genomic decoder. The loss is calculated similar to latent loss ($L_l$), between the $G_l$ and the output of genomic decoder $\hat{G_l}$.
    \vspace{-5pt}
    \[
    L_t = \lambda_1 D_{KL}(G_l, \hat{G_l}) + \lambda_2 \|G_l - \hat{G_l}\|^2
     \vspace{-5pt}
    \]
     \vspace{-5pt}
\end{itemize}
 \vspace{-5pt}
The final alignment loss function is the sum of the latent and translation losses:
\vspace{-7pt}
\[
L_{\text{alignment}} = L_l + L_t
\]
This loss ensures that the representations of pathology and genomic data are consistent and robust, allowing the model to effectively integrate both modalities for survival prediction.

\vspace{-5pt}

\section{Experiment and Results}
\vspace{-5pt}


\begin{table*}

    \centering
    \small
      \begin{minipage}{0.65\textwidth}
        \centering

    \begin{tabular}{c c c c c c c} \hline 
          Train&Test&  Methods&  BRCA&  LUAD& GBM& Mean \\ \hline 
         G &G&  Neural-Cox&  0.67 $\pm$ 0.088&  0.63 $\pm$ 0.019& 0.86 $\pm$ 0.036 & 0.72 \\ \hline 
         I &I&  MaxMIL& 0.57 $\pm$ 0.042  & 0.52 $\pm$ 0.021 & 0.66 $\pm$ 0.016 & 0.58\\ 
         I &I&  MeanMIL& 0.61 $\pm$0.086 & 0.58 $\pm$ 0.032 &  0.73$\pm$ 0.037 & 0.64\\ 
         I &I&  A-MIL& 0.61 $\pm$ 0.083  & 0.59$\pm$ 0.054 & 0.74 $\pm$ 0.042 & 0.64 \\ 
         I &I&  TransMIL&  0.59 $\pm$ 0.058&  0.58 $\pm$ 0.006& 0.79 $\pm$ 0.034 & 0.65 \\ \hline 
 I+G&I& SimL& 0.61 $\pm$ 0.085 & $0.60 \pm 0.015$& $0.74\pm0.014$ & 0.65 \\ 
 I+G&I& PathoGen-X& \textbf{0.67 $\pm$ 0.020}& \textbf{0.62 $\pm$ 0.008}&\textbf{0.81 $\pm$ 0.0023} & \textbf{0.70} \\ \hline
    \end{tabular}
    \caption{Survival prediction results (cross-validated on four-folds) on TCGA-BRCA, TCGA-LUAD, and TCGA-GBM datasets evaluated using C-index. The "Train" and "Test" columns indicate the modality used for training and testing, where "I" represents pathology images and "G" denotes genomic data. The best test results achieved using only imaging data are shown in \textbf{bold}.}
    \label{tab:results}
     \end{minipage}%
    \hfill
      \begin{minipage}{0.27\textwidth}
        \centering
\vspace{6mm} 
    \begin{tabular}{c c} \hline 
          Alignment Loss & BRCA  \\ \hline 
          $L_l$ & 0.64\\ \hline
          $L_t$ & 0.65 \\ \hline 
          $L_l + L_t$ & 0.67\\ \hline
   
    \end{tabular}
    \caption{Ablation study analyzing the impact of different alignment loss variations on the TCGA-BRCA dataset.}
    \label{tab:ablation_results}
    
     \end{minipage}%
     \vspace{-5mm}
\end{table*}

\subsection{Dataset}
We apply our method to three cancer survival datasets: TCGA-BRCA (n = 987), TCGA-LUAD (n = 509), and TCGA-GBM (n = 576). Each dataset contains WSIs and gene-expression data, along with labeled overall survival times (which represents the duration of time (in days) from the date of diagnosis (or start of treatment) until the patient's death or last follow-up) and right-censorship statuses. For our analysis, we selected 746 genes identified as prognostically important \cite{cosmic, krishna2024advancing}. From the gene expression data, we extracted FPKM-normalized RNA-seq data, which was then log-normalized. Model performance was evaluated using the concordance index (C-Index) to assess the model’s ability to rank pairs of individuals accurately based on predicted survival times. The pathology data is pre-processed using CLAM \cite{clam} through segmentation, patching, and feature extraction, with median blurring applied to smooth edges. Patches of size 256x256 were used, and features were extracted using a pretrained ResNet50 model, converting each patch into a 1024-dimensional feature vector.

\vspace{-8pt}

\subsection{Implementation details}
We compared our method with the following approaches: (1) Mean-MIL and Max-MIL, (2) Attention-MIL \cite{attentionmil}, (3) TransMIL \cite{transmil}, and (4) a similarity learning (SimL) approach~\cite{ding2023pathology}, where genomic and pathology embeddings are mapped to a shared latent space through unsupervised pre-training. Subsequently, the pathology embeddings were fine-tuned for the survival prediction task. Additionally, we implemented a genomic-based survival prediction model using a neural Cox model \cite{neuralcox} to demonstrate that genomic data is a stronger predictive modality than pathology images. 

Our model was developed using the PyTorch framework and trained on a Nvidia A100 GPU.  Adam optimizer was used, with a learning rate of 0.001 and L2 regularization of 0.1. Training was conducted over 12 epochs with a batch size of 128, using 4-fold cross-validation to evaluate performance, with a train-validation split ratio of 3:1.

\vspace{-4pt}

\subsection{Results}
\begin{figure}[h]
    \centering
    \vspace{-1pt}
    \includegraphics[scale=0.36] {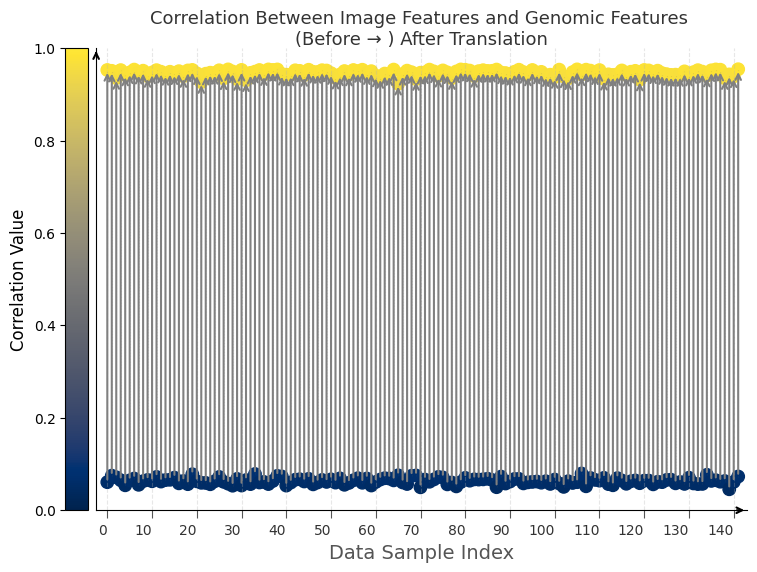}
    \vspace{-6pt}
    \caption{Visualization of the substantial  improvement in correlation between the image features with genomic features after feature translation for GBM dataset samples. }
    \label{fig:corr}
    \vspace{-4mm}
\vspace{-10pt}
\end{figure}

As shown in Table \ref{tab:results}, our method outperforms all competing approaches that rely on pathology images for testing. It surpasses the second-best method by c-index margins of 0.06, 0.03, and 0.02 for BRCA, LUAD, and GBM datasets, respectively, with an average improvement of 0.05. Additionally, the table indicates that methods based on genomic data achieve better performance than those relying on pathology images, highlighting genomic data is a stronger predictive modality. Our objective was to align the image-based model's results closely with those of the genomic model, and we have successfully accomplished this, as evident from the results. Figure \ref{fig:corr} illustrates an increased correlation between image and genomic features following the translation process, indicating that the model is aligning the features as intended. We also plotted Kaplan Meier (KM) survival curves in Figure \ref{fig:surv_curves} obtained using our method, that indicated significant disparity in the survival outcomes between high and low risk groups with high statistical significance (all $p<0.01$). Finally in Table \ref{tab:ablation_results}, we conduct an  ablation study on our framework to analyze the effect of each loss on our results, which indicates that both losses together contribute to the high performance. 
\vspace{-4pt}
\vspace{-10pt}
\section{Discussion and Conclusion}
\vspace{-3pt}
Our results indicate that our model outperforms previous image-based models by using feature alignment with genomic data during training. While the unsupervised pre-training approaches demonstrates improved performance with large datasets, as noted in \cite{ding2023pathology}, it also introduces some redundancy that can hinder its effectiveness. Our results on SimL model suggest that it is not an optimal method for task-specific objectives, especially when data is limited. The pathology features derived from our pre-trained model can also be utilized for tasks more closely related to survival, such as treatment planning and drug response prediction. Furthermore, the alignment and translation methods we have implemented can be adapted for other downstream tasks, e.g., classification.

\vspace{-4mm}
\section{Compliance with Ethical Standards}
\label{sec:ethics}
This research study was conducted retrospectively using human subject data made available in open access by \cite{tcga}. Ethical approval was not required as confirmed by the license attached with the open access data.
\section{Acknowledgments}
\label{sec:acknowledgments}
The results of this study are based on the data collected from the public TCGA Research Network \href{https://www.cancer.gov/tcga}.

\bibliographystyle{IEEEbib}
\bibliography{strings,refs}

\begin{thebibliography}{10}

\bibitem{cherian20212021}
Nikhil Cherian~Kurian, Amit Sethi, Anil Reddy~Konduru, Abhishek Mahajan, and Swapnil~Ulhas Rane,
\newblock ``A 2021 update on cancer image analytics with deep learning,''
\newblock {\em Wiley Interdisciplinary Reviews: Data Mining and Knowledge Discovery}, vol. 11, no. 4, pp. e1410, 2021.

\bibitem{verghese2023multiscale}
Gregory Verghese, Mengyuan Li, and Liu et. al,
\newblock ``Multiscale deep learning framework captures systemic immune features in lymph nodes predictive of triple negative breast cancer outcome in large-scale studies,''
\newblock {\em The Journal of pathology}, vol. 260, no. 4, pp. 376--389, 2023.

\bibitem{ding2023pathology}
Kexin Ding, Mu~Zhou, Dimitris~N Metaxas, and Shaoting Zhang,
\newblock ``Pathology-and-genomics multimodal transformer for survival outcome prediction,''
\newblock in {\em International Conference on Medical Image Computing and Computer-Assisted Intervention}. Springer, 2023, pp. 622--631.

\bibitem{transmil}
Zhuchen Shao, Hao Bian, Yang Chen, Yifeng Wang, Jian Zhang, Xiangyang Ji, et~al.,
\newblock ``Transmil: Transformer based correlated multiple instance learning for whole slide image classification,''
\newblock {\em Advances in neural information processing systems}, vol. 34, pp. 2136--2147, 2021.

\bibitem{attentionmil}
Maximilian Ilse, Jakub Tomczak, and Max Welling,
\newblock ``Attention-based deep multiple instance learning,''
\newblock in {\em International conference on machine learning}. PMLR, 2018, pp. 2127--2136.

\bibitem{clam}
Ming~Y Lu, Drew~FK Williamson, Tiffany~Y Chen, Richard~J Chen, Matteo Barbieri, and Faisal Mahmood,
\newblock ``Data-efficient and weakly supervised computational pathology on whole-slide images,''
\newblock {\em Nature biomedical engineering}, vol. 5, no. 6, pp. 555--570, 2021.

\bibitem{zhou2023cross}
Fengtao Zhou and Hao Chen,
\newblock ``Cross-modal translation and alignment for survival analysis,''
\newblock in {\em Proceedings of the IEEE/CVF International Conference on Computer Vision}, 2023, pp. 21485--21494.

\bibitem{tcga}
Kyle et.~al. Chang,
\newblock ``The cancer genome atlas pan-cancer analysis project,''
\newblock {\em Nature Genetics}, 2013.

\bibitem{mobadersany2018predicting}
Pooya Mobadersany, Safoora Yousefi, Mohamed Amgad, David~A Gutman, Jill~S Barnholtz-Sloan, Jos{\'e}~E Vel{\'a}zquez~Vega, Daniel~J Brat, and Lee~AD Cooper,
\newblock ``Predicting cancer outcomes from histology and genomics using convolutional networks,''
\newblock {\em Proceedings of the National Academy of Sciences}, vol. 115, no. 13, pp. E2970--E2979, 2018.

\bibitem{zheng2022multi}
Hanci Zheng, Zongying Lin, Qizheng Zhou, Xingchen Peng, Jianghong Xiao, Chen Zu, Zhengyang Jiao, and Yan Wang,
\newblock ``Multi-transsp: Multimodal transformer for survival prediction of nasopharyngeal carcinoma patients,''
\newblock in {\em International Conference on Medical Image Computing and Computer-Assisted Intervention}. Springer, 2022, pp. 234--243.

\bibitem{long2024mugi}
Lifan Long, Jiaqi Cui, Pinxian Zeng, Yilun Li, Yuanjun Liu, and Yan Wang,
\newblock ``Mugi: Multi-granularity interactions of heterogeneous biomedical data for survival prediction,''
\newblock in {\em International Conference on Medical Image Computing and Computer-Assisted Intervention}. Springer, 2024, pp. 490--500.

\bibitem{attention}
A~Vaswani,
\newblock ``Attention is all you need,''
\newblock {\em Advances in Neural Information Processing Systems}, 2017.

\bibitem{cox}
H{\aa}vard Kvamme, {\O}rnulf Borgan, and Ida Scheel,
\newblock ``Time-to-event prediction with neural networks and cox regression,''
\newblock {\em Journal of machine learning research}, vol. 20, no. 129, pp. 1--30, 2019.

\bibitem{cosmic}
Zbyslaw Sondka, Nidhi~Bindal Dhir, Denise Carvalho-Silva, Steven Jupe, Madhumita, Karen McLaren, Mike Starkey, Sari Ward, Jennifer Wilding, Madiha Ahmed, et~al.,
\newblock ``Cosmic: a curated database of somatic variants and clinical data for cancer,''
\newblock {\em Nucleic Acids Research}, vol. 52, no. D1, pp. D1210--D1217, 2024.

\bibitem{krishna2024advancing}
Akhila Krishna, Ravi~Kant Gupta, Pranav Jeevan, and Amit Sethi,
\newblock ``Advancing gene selection in oncology: A fusion of deep learning and sparsity for precision gene selection,''
\newblock {\em arXiv preprint arXiv:2403.01927}, 2024.

\bibitem{neuralcox}
Xinliang Zhu, Jiawen Yao, and Junzhou Huang,
\newblock ``Deep convolutional neural network for survival analysis with pathological images,''
\newblock in {\em 2016 IEEE international conference on bioinformatics and biomedicine (BIBM)}. IEEE, 2016, pp. 544--547.

\end{thebibliography}

\end{document}